\documentclass[tradiabstract,rnote]{aa}
\usepackage{graphicx,amssymb}
\usepackage{array}
\usepackage{natbib}
\usepackage{caption}
\usepackage{amsmath}
\usepackage{bm}
\usepackage{multirow}
\usepackage{geometry}
\usepackage{lscape}
\usepackage[breaklinks, dvips]{hyperref}
\usepackage{subcaption}
\usepackage[varg]{txfonts}
\begin{document}

\title{Total infrared luminosity estimation from local galaxies in AKARI all sky survey}

\author{A.~Solarz\inst{1}
\and T.~T.~Takeuchi\inst{2}
  \and A.~Pollo\inst{1,3}
}

\institute{National Center for Nuclear Research, ul. Hoża 69, 00-681 Warsaw, Poland\\ \email{aleksandra.solarz@ncbj.gov.pl}
         \and
        Division of Particle and Astrophysical Science, Nagoya University, Furo-cho, Chikusa-ku, Nagoya 464-8602, Japan
         \and
         The Astronomical Observatory of the Jagiellonian University, ul.\ Orla 171, 30-244 Kraków, Poland}
\date{Received <date>/ Accepted <date>}
\abstract{We aim to use the a new and improved version of AKARI all sky survey catalogue of far-infrared sources to recalibrate the formula to derive the total infrared luminosity. We cross-match the faint source catalogue (FSC) of IRAS with the new AKARI-FIS and obtained a sample of 2430 objects. Then we calculate the total infrared (TIR) luminosity $L_{\textrm{TIR}}$ from the Sanders at al. (1996) formula and compare it with total infrared luminosity from AKARI FIS bands to obtain new coefficients for the general relation to convert FIR luminosity from AKARI bands to the TIR luminosity.}


\maketitle 
\section{Introduction}
The question of how and when galaxies have formed their stellar content is paramount when addressing the topic of formation and evolution of galaxies. 
The baryonic gas contained within a dark matter halo goes through a cooling process, which results in a decrease of pressure. As a result, it starts to flow into the centre of the halos' potential well, creating an excess of density. When it exceeds the density of the dark matter, the gas starts to act under its own gravity and eventually collapses. 
This process ultimately leads to the creation of a dense cloud of molecular gas, which will serve as a 'nursery' for the star formation (e.g. \citealt{williams}).

The process of formation of stars in galaxies is always accompanied by the dust production (e.g. \citealt{schulz}, \citealt{stahler}); however, the exact physics of this process is still a subject of investigation (e.g. \citealt{ttt05}).
One thing is certain, though: the dust particles in galaxies absorb the ultra-violet (UV) radiation emitted by young, hot stars, and re-emit it in the far infrared (FIR). Therefore, the observations at these wavelengths are in tight correlation with the amount of the dust contained in a galaxy and with its stellar content properties.
Majority of the star forming regions within a galaxy are hidden in dense gas clouds (e.g. \citealt{sasiadka}), which makes them impossible to observe in the UV.
This fact is also confirmed by analyses of the evolution of the luminosity function (the dependence of the number of stars/galaxies on their absolute magnitude) both in FIR and UV \citep{ttt05a}, with the former one being far stronger than the latter, which means that the amount of the hidden star formation zones grows proportionally with redshift (for $z<1$).
All those reasons make the infrared observations of the Universe crucial for understanding the star formation history of the Universe. 

One of the ways to quantify star-formation is to measure the total infrared luminosity ($L_{\textrm{TIR}}$), as the two are tightly correlated (e.g. \citealt{helou88}, \citealt{sanders96}, \citealt{dale01}, \citealt{dale02}, \citealt{draine07}, \citealt{takeuchi10}).
Therefore, by using the emission measured throughout different IR passbands it is possible to recover the information about the star-formation activity of galaxies.

The purpose behind the launch of AKARI satellite was to make all-sky surveys at infrared wavelengths with better sensitivity, spatial resolution and wavelength coverage than that of its predecessor: pioneering IRAS \citep{soi}. Other past IR satellites, ISO \citep{kess} and Spitzer Space Telescope \citep{wer}, though having better  capabilities than IRAS, were not designed for all-sky surveys.
The AKARI satellite was launched by JAXA’s MV8 vehicle on
February 22, 2006, and one of its main goals was to perform and all-sky survey at far-infrared (FIR) wavelengths. For this purpose a dedicated FIS camera \citet{kawada07} was used. It covered the wavelength range between 50 and 180~$\mathrm{\mu}$m through exposure of 4 FIR filters: \textit{N60} centred at 65~$\mathrm{\mu}$m, WIDE-S centred at 90~$\mathrm{\mu}$m, WIDE-L centred at 140~$\mathrm{\mu}$m and \textit{N160} centred at 160~$\mathrm{\mu}$m. 

In this work, we use new and secure data from AKARI all sky survey to present a recalibrated formula to recover $L_{\textrm{TIR}}$.
We adopt a cosmological model with $\Omega_{m}$ = 0.3, and $\Omega_{\Lambda}$ = 0.7, $H_{0}=70$~km~$\mathrm{s^{-1}}$~$\mathrm{Mpc^{-1}}$ and define $L_{\nu}$~$[\mathrm{erg}$~$\mathrm{s^{-1}}$~$\mathrm{Hz^{-1}}]$ as the luminosity per unit frequency at a frequency $\nu = c/\lambda$ (c: the speed of light) throughout this paper.
\section{Data}

The newest version of the AKARI all sky survey FIS catalogue consists of 950~365 sources, out of which 410~623 (43\%) posses measurements in all four passbands.
The point spread function of AKARI FIR bands is reported to be $37\pm1"$ for \textit{N60} band, $39\pm1"$ for WIDE-S band, $58\pm3"$ for WIDE-L band and $61\pm4"$ for \textit{N160} band (\citealt{kawada07}).
To ensure a good quality of photometric measurements we choose those objects that have been scanned by AKARI at least three times and posses the flux quality indicator FQUAL=3 at 90~$\mu$m.
The depth of the catalogue at 90~$\mathrm{\mu}$m filter where we consider only sources with signal to noise ratios above 5, reaches $\sim 0.2$ [Jy].

\begin{figure*}
  \centering
  \includegraphics[width=.65\linewidth]{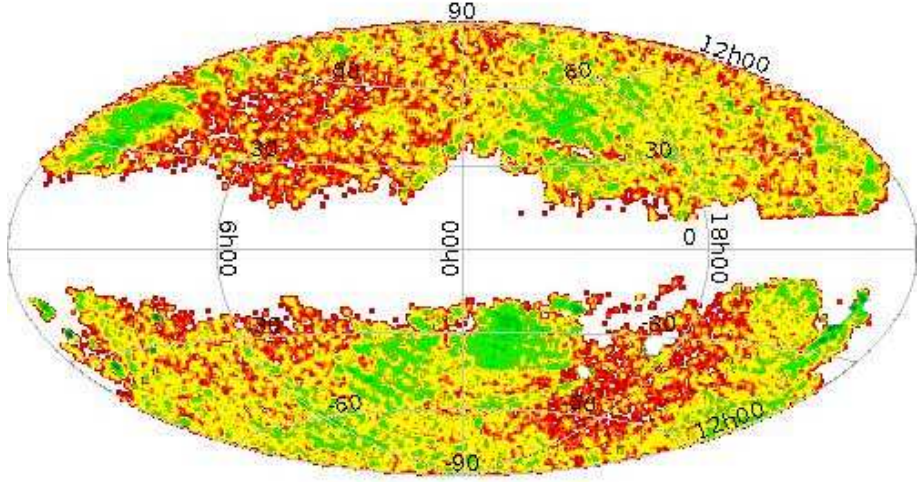}
  \caption{The sky distribution of the AKARI FIS sources from the new catalogue with areas of high 100~$\mathrm{\mu}$m emission removed.}
\label{allskypos}
\end{figure*}

Following \citep{pollo10} we select objects lying in regions associated with low Galactic emission at 100~$\mathrm{\mu}$m.  
To obtain the 100~$\mathrm{\mu}$m emission from the Milky Way we used the SFD maps \citep{schlegel98}.
We have set the threshold to 5~MJy~$\mathrm{sr^{-1}}$, which leaves us with a catalogue composed of 61~576 (6\%) sources. 
The distribution of those sources on the sky is shown in Fig.~\ref{allskypos}.
\subsection{Matching with IRAS FSC catalogue}
In order to provide reliable estimates of the total infrared luminosities, a robust catalogue of FIR sources confirmed as galaxies is needed. 
To this aim, we cross-matched the AKARI FIS bright source
catalogue ver. 2 with IRAS faint source catalogue (FSC; \citealt{moshir92}). The FSC contains data for 173~044 point sources in unconfused regions with flux densities typically greater than 0.2~Jy at 12, 25, and 60~$\mathrm{\mu}$m and greater than 0.4~Jy at 100~$\mathrm{\mu}$m.
We searched for IRAS FSC counterparts within AKARI BSCv2 in a radius of 20 arcsec, corresponding to the position uncertainty of the IRAS FSC. When a source had multiple counterparts detected in IRAS FSC catalogue, the closest matching object was selected as a counterpart. The total number of crossmatched sources was $2430$ and there are measurements obtained
by \textit{IRAS} satellite in all its passbands for all those objects.

\begin{figure}
  \centering
  \includegraphics[width=.69\linewidth]{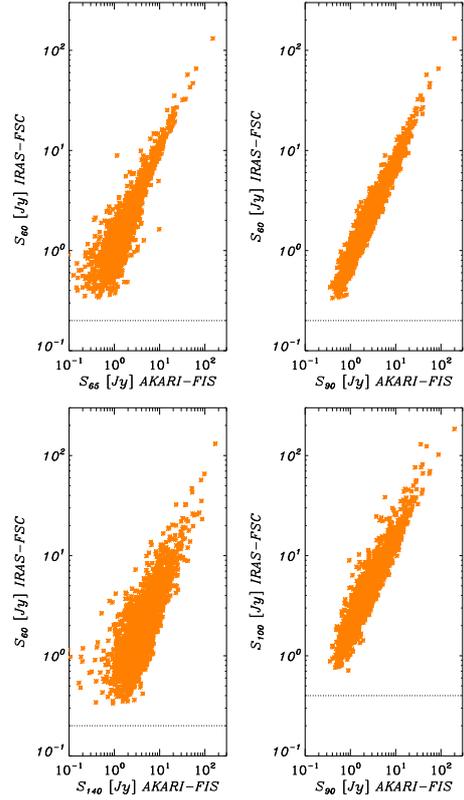}
  \caption{Comparison between the AKARI FIS and IRAS FSC flux densities. Upper-left, upper-right, and lower-left panels present comparisons of the IRAS 60~$\mathrm{\mu}$m with AKARI 90~$\mathrm{\mu}$m, 65~$\mathrm{\mu}$m, and 140~$\mathrm{\mu}$m flux densities of the AKARI-IRAS correlated sample. Lower-right panel shows a comparison of the IRAS 100~$\mathrm{\mu}$m with the AKARI 90~$\mathrm{\mu}$m flux densities. The horizontal dashed lines in these panels represent the flux density limit of the IRAS FSC.}
\label{akariiras}
\end{figure}

We compared the AKARI and IRAS flux densities to examine
our sample selection to see which selection controls the sample selection. The relation is presented in Fig.~\ref{akariiras}.
 The AKARI FIS selection of sources is based on WIDE-S measurements and after the crossmatching procedure we found that the 90~$\mathrm{\mu}$m flux density limit of the final source sample is $\sim$~0.3 Jy.


\section{Total infrared luminosity calibration}

To measure the total infrared luminosity ($L_{\textrm{TIR}}$) we used a classical formula measuring luminosity between\\ $\lambda=8-1000$~$\mu$m presented by \citet{sanders96} for \textit{IRAS} bands (centred at 12~$\mathrm{\mu}$m, 25~$\mathrm{\mu}$m, 60~$\mathrm{\mu}$m and 100~$\mathrm{\mu}$m):
\begin{equation}
\begin{split}
L_{\mathrm{TIR}} [L_{\odot}]=&4.93\times 10^{-22}[13.48L_{\nu}(12~\mu\mbox{m})+5.16L_{\nu}(25\mu\mbox{m})+\\
 &\quad+2.58L_{\nu}(60\mu\mbox{m})+L_{\nu}(100\mu\mbox{m})].
\end{split}
\label{sm}
\end{equation}

To obtain specific luminosities per unit frequency ($L_{\nu}$) we first measured the luminosity distance to each source as 
\begin{eqnarray}
D_{L}=&(1+z)D_{H}\int^{z}_{0}dz'/E(z'),
\end{eqnarray}where
\begin{eqnarray}
 E(z)=&\sqrt{\Omega_{M}(1+z)^{3}+\Omega_{k}(1+z)^{2}+\Omega_{\Lambda}),}
\end{eqnarray}
and
\begin{eqnarray}
D_{H}=&c/H_{0,}
\end{eqnarray}

is the Hubble distance. Then, to calculate specific luminosities we used $D_{L_{\nu}}=\sqrt{L_{\nu}/4\pi S_{\nu}}$, where $S_{\nu}$ is the flux density measured for each source through a specific filter.
To obtain the relationships between $L_{\textrm{TIR}}$ and specific AKARI FIS luminosities we fitted a following model:\\ $y=a*x+b$, where $y=\mathrm{log}( L_{\textrm{TIR}})$ and $x=\mathrm{log}(L_{AKARI-\mbox{ FIS filter}})$.

\begin{figure*}
\begin{subfigure}{.5\textwidth}
  \centering
  \includegraphics[width=.95\linewidth]{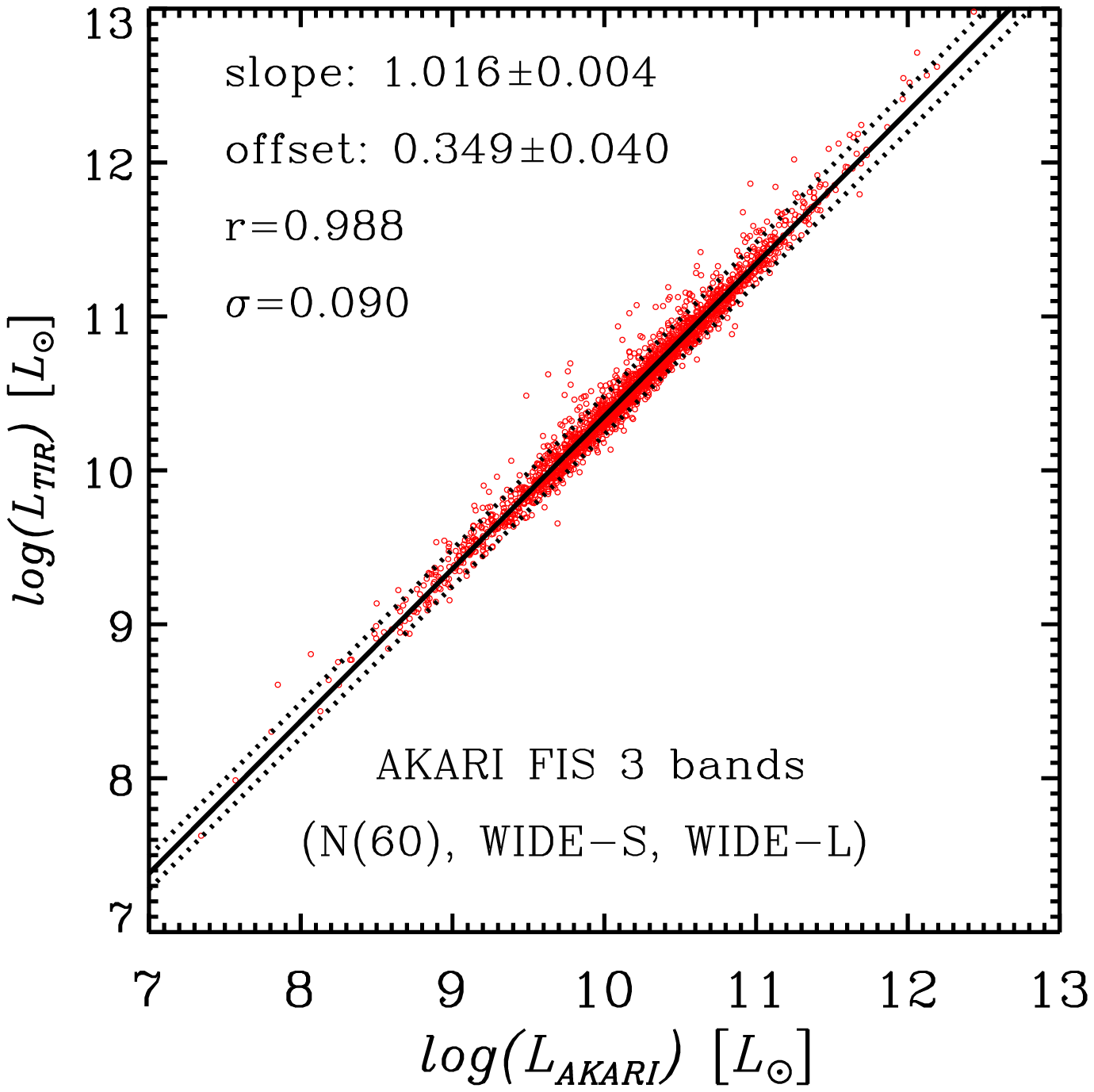}

\end{subfigure}%
\begin{subfigure}{.5\textwidth}
  \centering
  \includegraphics[width=.95\linewidth]{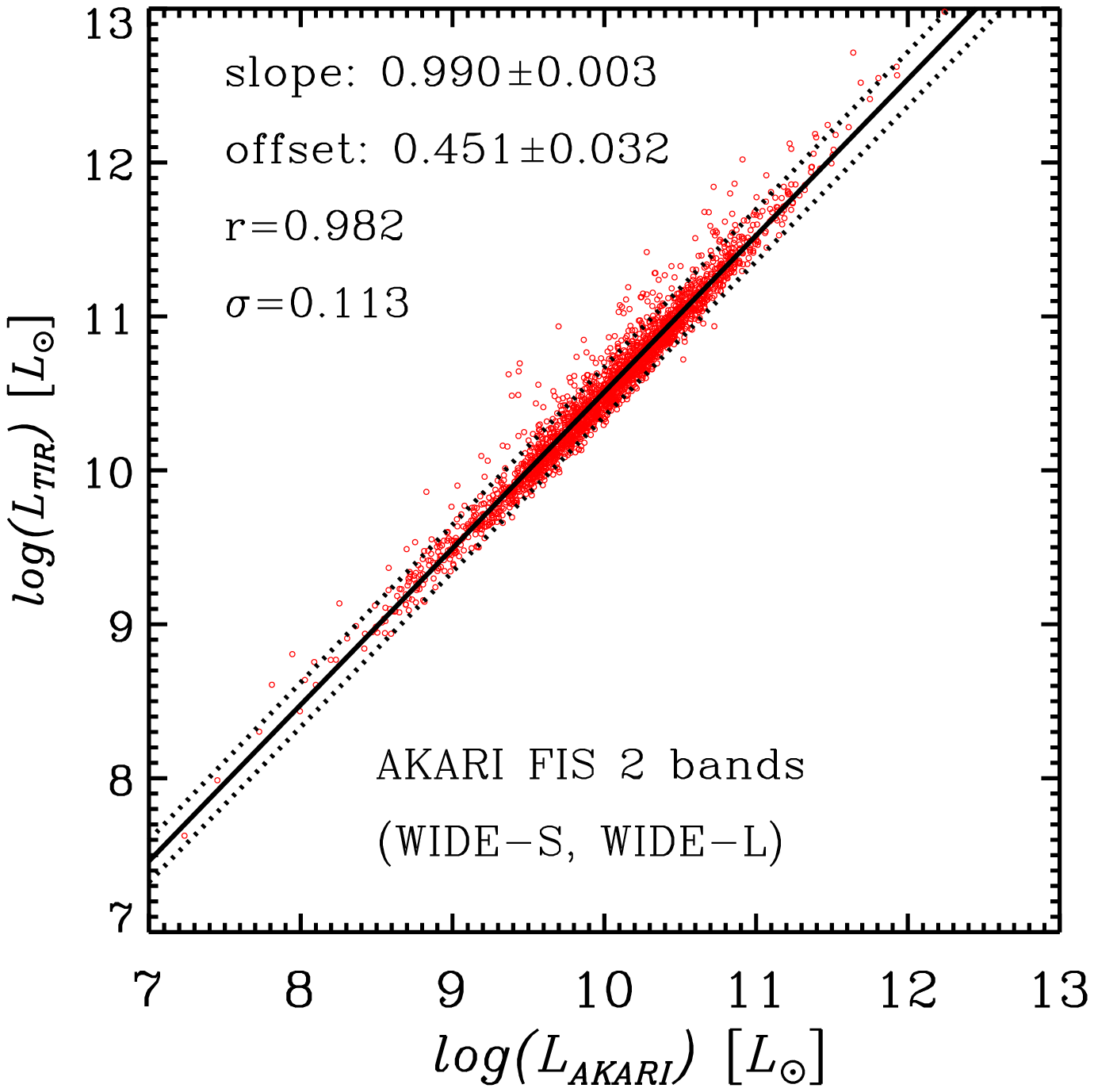}

\end{subfigure}
\caption{ Relation between $L_{AKARI}$ estimated from three (left panel) and two (right panel) AKARI FIS bands and $L_{\textrm{TIR}}$ estimated from \citet{sanders96}. The solid lines present the best least-square fits to the data, and the dashed lines mark the 95\% prediction interval of the linear regression.}
\label{lakari}
\end{figure*}

Following \citet{hirashita08} and \citet{takeuchi10} we took advantage of the continuous wavelength coverage of AKARI FIS bands (from $\sim50$~$\mathrm{\mu}$m to $\sim160$~$\mathrm{\mu}$m) and used their definition of IR flux by taking a sum of flux densities multiplied by their bandwidths:

\begin{equation}
\begin{split}
L_{AKARI}^{\mathrm{3 bands}}&=\Delta \nu(N\mathit{60})L_{\nu}(65~\mathrm{\mu} m)+\\&\quad+\Delta \nu(\mbox{WIDE-S})L_{\nu}(90~\mathrm{\mu} m)+\\&\quad+\Delta \nu(\mbox{WIDE-L})L_{\nu}(140~\mathrm{\mu} m),
\end{split}
\end{equation}
where\\
$\Delta \nu(N\mathit{60})=1.58 \times 10^{12}$[Hz],\\
$\Delta \nu(\mbox{WIDE-S})=1.47 \times 10^{12}$[Hz],\\
$\Delta \nu(\mbox{WIDE-L})=0.831\times 10^{12}$[Hz].\\

As \citet{takeuchi10} reported, N60 filter sensitivity is lower than that of the wide bands, and therefore defined IR luminosity by using only WIDE-S and WIDE-L bands:

\begin{equation}
\begin{split}
L_{AKARI}^{\mathrm{2 bands}}&=\Delta \nu(\mathrm{WIDE-S})L_{\nu}(90~\mathrm{\mu} m)+\\ \quad &+\Delta \nu(\mathrm{WIDE-L})L_{\nu}(140~\mathrm{\mu} m).
\end{split}
\end{equation}
\\
In Fig.~\ref{lakari} we present correlations between $L_{\mathrm{TIR}}$ and both definitions of $L_{AKARI}$.
We found the relations (shown in Fig.~\ref{lakari}) as follows:
\begin{equation}
\begin{split}
\mathrm{log}(L_{\mathrm{TIR}})=&(1.016\pm0.004)\mathrm{log}(L_{AKARI}^{\mathrm{3 bands}})+\\ \quad &+(0.349\pm0.040)
\end{split}
\end{equation}
with correlation coefficient $r=0.988$, and 
\begin{equation}
\begin{split}
\mathrm{log}(L_{\mathrm{TIR}})=&(0.990\pm0.003)\mathrm{log}(L_{AKARI}^{\mathrm{2 bands}})+\\ \quad &+(0.451\pm0.032)
\end{split}
\end{equation}
with $r=0.982$. The envelopes delineated by the dashed lines are 95\% prediction interval of the linear regression. The scatter around the two best fits is approximately equal, with $\sigma=0.11$ for $L_{AKARI}^{\mathrm{2 bands}}$ and $\sigma=0.09$ for $L_{AKARI}^{\mathrm{3 bands}}$.

\section{Conclusions and summary}
In this work we used the new release of the AKARI all sky survey catalogue performed at far-infrared and covering the wavelength range from 50 to 180 $\mathrm{\mu}$m to recalibrate the formula to obtain total infrared luminosities.
We find very tight correlations between the luminosity estimates from AKARI FIS catalogue and luminosity estimates from IRAS FSC catalogue. 
We do not find any significant enhancement of the correlations when only two wide filters are used instead of three AKARI-FIS filter combination ($N60$, $\mbox{WIDE-S}$, $\mbox{WIDE-L}$). 

\begin{acknowledgements}
This work is based on observations with AKARI, a JAXA project with the participation of ESA.
We thank I. Yamamura and K. Murata (ISAS) for providing useful comments which helped to improve this manuscript.
AS was supported by the National Science Centre grant UMO-2015/16/S/ST9/00438.
TTT has been supported by the Grant-in- Aid for the Scientific Research Fund (24111707), commissioned by the
Ministry of Education, Culture, Sports, Science and Technology (MEXT) of Japan, and for the JSPS Strategic Young Researcher Overseas Visits Program for Accelerating Brain Circulation, “Construction of a Global 7 Platform for the Study of Sustainable Humanosphere.”
AP has been supported by the National Science Centre grant UMO-2012/07/B/ST9/04425.
The authors would like to thank the anonymous referee for comments and recommendations that have improved this work.
\end{acknowledgements}

\bibliographystyle{aa}
\bibliography{solarz.bib}
\end{document}